\documentclass[prd,aps,twocolumn,amsmath,amssymb,nofootinbib,preprintnumbers]
{revtex4}

\usepackage{graphicx}
\usepackage{dcolumn}
\usepackage{bm}
\usepackage{amsmath}
\usepackage{amsfonts}

\def\ls{\mathrel{\lower4pt\vbox{\lineskip=0pt\baselineskip=0pt
           \hbox{$<$}\hbox{$\sim$}}}}
\def\gs{\mathrel{\lower4pt\vbox{\lineskip=0pt\baselineskip=0pt
           \hbox{$>$}\hbox{$\sim$}}}}
\def\drawbox#1#2{\hrule height#2pt

\hbox{\vrule width#2pt height#1pt \kern#1pt
              \vrule width#2pt}
              \hrule height#2pt}

\def\Asym#1#2{\vcenter{\vbox{\drawbox{#1}{#2}
              \kern-#2pt       
              \drawbox{#1}{#2}}}}

\newcommand{\beq}{\begin{equation}}
\newcommand{\eeq}{\end{equation}}

\begin{document}

%
\title{Attraction towards an inflection point inflation}

\author{Rouzbeh Allahverdi$^{1}$}
\author{Bhaskar Dutta$^{2}$}
\author{Anupam Mazumdar$^{3,4}$}

\affiliation{$^{1}$~Department of Physics \& Astronomy, University of New Mexico, Albuquerque, NM 87131, USA \\ $^{2}$~Department of Physics, Texas A\&M
University, College Station, TX 77843-4242, USA\\
$^{3}$~Physics Department, Lancaster University, Lancaster, LA1 4YB, UK\\
$^{4}$~Niels Bohr Institute, Blegdamsvej-17, Copenhagen-2100, Denmark}


\begin{abstract}
Many models of high energy physics possess metastable vacua. It is conceivable that the universe can get trapped in such a false vacuum, irrespective of its origin and prior history, at an earlier stage during its evolution. The ensuing false vacuum inflation results in a cold and empty universe and has a generic graceful exit problem. We show that an inflection point inflation along the flat directions of the Minimal Supersymmetric Standard Model (MSSM) can resolve this graceful exit problem by inflating the bubble which nucleate out of a false vacuum. The important point is that the initial condition for an MSSM inflation can be naturally realized, due to an attractor behavior towards the inflection point. We investigate these issues in detail and also present an example where metastable vacua, hence the false vacuum inflation, can happen within the MSSM.

\end{abstract}
MIFP-08-14, June, 2008
\maketitle
\section{Introduction}

Metastable vacua can be found in supersymmetric and non-supersymmetric theories.
In particular, they are believed to be common in string theory, known as string landscape~\cite{Douglas}. Irrespective of the very initial conditions of the universe, it is likely that as the universe expanded it found itself trapped inside a false vacuum at some point. A false vacuum would lead to a homogeneous and an isotropic universe by virtue of inflation~\cite{Guth}. The universe would eventually tunnel to either a lower nearest vacuum or to the lowest available ground state. In either case graceful exit from a false vacuum inflation remains a challenge, i.e. to end up with an observable universe consisting of the Standard Model (SM) baryons and cold dark matter and with the right amplitude of density perturbations~\cite{WMAP}.

One can imagine tunneling from a false vacuum followed by a phase of slow roll inflation. This secondary phase of inflation would inflate the bubble to the size of an observable patch and generated the desired density perturbations. Also, (dark) matter would be created during the reheating phase with an abundance which is in agreement with cosmological observations~\cite{WMAP,BBN}.

The Minimal Supersymmetric Standard Model (MSSM) provides all the necessary ingredients for a successful slow roll inflation~\cite{AEGM,AEGJM,AKM,AJM}. The inflaton candidates are two flat directions consisting of squarks and sleptons: $udd$ (where $u$ and $d$ denote right-handed up type and down type squarks respectively) and $LLe$ (where $L$ denote left-handed sleptons and $e$ denotes a right handed charged slepton). There is also a candidate with a minimal extension of the Standard Model (SM) gauge group; the $N H_u L$ flat direction, where $N$ is a right-handed sneutrino and $H_u$ is the Higgs which gives mass to up-type quarks~\cite{AKM}. The supersymmetric flat directions, which are classified by {\it gauge invariant} combinations of superfields, have a vanishing potential in the limit of an exact supersymmetry~\footnote{MSSM flat directions have many applications, for example Affleck-Dine baryogenesis~\cite{DRT}, curvaton mechanism~\cite{Curvaton}, reheating and thermalization~\cite{Ours}, and generation of primordial magnetic fields~\cite{Asko}.}. However various contributions such as soft supersymmetry breaking terms and higher order superpotential terms lift the flatness of the potential~\cite{MSSM-REV}. MSSM inflation has remarkable properties which we summarize below:

\begin{itemize}

\item{There is no singlet field (with ad-hoc mass and couplings) involved in order to explain inflation. The inflaton candidates are made up of fields which are charged under the SM gauge symmetry, their couplings are known, and their masses are well motivated by the requirement of a low scale supersymmetry. It guarantees reheating to the SM degrees of freedom~\cite{AEGM,AEGJM}. Furthermore the parameter space of the model is compatible with thermal supersymmetric dark matter~\cite{ADM,ADM2}.}

\item{Inflation occurs at sub-Planckian VEVs and with large number of e-foldings. The flatness of the potential also allows a phase of eternal inflation~\cite{AEGM,AEGM}. The parameter space allows a wide range of scalar spectral index which matches the observed amplitude of the perturbations and the spectral tilt. In particular spectral index can vary within a wide range $0.92 \leq n_s\leq 1.0$ without affecting much the amplitude of the perturbations~\cite{AEGJM,ADM,LK}.}

\item{Inflation happens at a low scale which guarantees that the model is insensitive to
possible trans-Planckian corrections, and also supergravity corrections~\cite{AEGJM}. The
low scale also helps solving the moduli and the gravitino problems~\cite{AEGJM}.}

\end{itemize}

Given these remarkable properties a question remains as how natural the initial conditions for MSSM inflation are. Since the scale of inflation is very low, i.e. $H_{\rm MSSM} \sim {\cal O}(1)$~GeV~\cite{AEGM,AEGJM}, it is considered to be a challenging task to obtain the right initial conditions for inflation to occur. However as we shall show in this paper this issue is addressed rather naturally if the universe undergoes a phase of false vacuum inflation~\footnote{Ref.~\cite{AFM} discusses how the initial conditions can be set up within a string landscape.}. The key point is that MSSM inflation happens near a point of inflection. Interestingly, we find that the inflection point has as an attractor behavior in a background of false vacuum inflation. This implies that at late times the inflaton will end up near the point of inflection largely independently of its initial conditions. In a false vacuum, the bulk of space will undergo inflation forever. Hence most of the bubbles that nucleate after a long enough time would have the right initial conditions for MSSM inflation. This second phase of inflation begins when the flat direction inflaton dominates the energy density inside the bubble, and eventually leads to an emergence of the observable universe. We also show that it is possible to obtain metastable vacua along some other MSSM flat direction(s), which are orthogonal to the inflaton candidates, without adding any extra fields. This would lead to an intriguing possibility that the entire episode could happen within MSSM.

We begin this paper with a brief review of MSSM inflation. Then we move on to exploring the MSSM parameter space in the light of new WMAP data. Next we describe how initial conditions can be set during a phase of false vacuum inflation as a virtue of the attractive behavior of the inflection point. We then show that MSSM can allow false vacua at large VEVs. Finally we close with a discussion and conclusions.


\section{A brief on MSSM inflation}

Let us recapitulate the main features of MSSM flat direction inflation~\cite{AEGM,AEGJM}. The $udd$ and $LLe$ flat directions are lifted by higher order superpotential terms of the following form~\cite{GKM}
\beq \label{supot}
W \supset {\lambda \over 6}{\Phi^6 \over M^3_{\rm P}}\, ,
\eeq
where $\lambda \sim {\cal O}(1)$. The scalar component of $\Phi$ superfield, denoted by $\phi$, is given by
\beq \label{infl}
\phi = {{u} + {d} + {d} \over \sqrt{3}} ~ ~ ~ , ~ ~ ~ \phi = {{L} + {L} + {e} \over \sqrt{3}},
\eeq
for the $udd$ and $LLe$ flat directions respectively.

After minimizing the potential along the angular direction $\theta$, we can situate the real part of $\phi$ by rotating it to the corresponding angles $\theta_{\rm min}$. The scalar potential is then found to be~\cite{AEGM,AEGJM}
\beq \label{scpot}
V(\phi) = {1\over2} m^2_\phi\, \phi^2 - A {\lambda\phi^6 \over 6\,M^{6}_{\rm P}} + \lambda^2
{{\phi}^{10} \over M^{6}_{\rm P}}\,,
\eeq
where $m_\phi$ and $A$ are the soft breaking mass and the $A$-term respectively ($A$ is a positive quantity since its phase is absorbed by a redefinition of $\theta$ during the process).


\subsection{Inflation near a point of inflection}

Provided that
\beq \label{dev}
{A^2 \over 40 m^2_{\phi}} \equiv 1 + 4 \alpha^2\, ,
\eeq
where $\alpha^2 \ll 1$, there exists a point of inflection in $V(\phi)$
\begin{eqnarray}
&&\phi_0 = \left({m_\phi M^{3}_{\rm P}\over \lambda \sqrt{10}}\right)^{1/4} + {\cal O}(\alpha^2) \, , \label{infvev} \\
&&\, \nonumber \\
&&V^{\prime \prime}(\phi_0) = 0 \, , \label{2nd}
\end{eqnarray}
at which
\begin{eqnarray}
\label{pot}
&&V(\phi_0) = \frac{4}{15}m_{\phi}^2\phi_0^2 + {\cal O}(\alpha^2) \, , \\
\label{1st}
&&V'(\phi_0) = 4 \alpha^2 m^2_{\phi} \phi_0 \, + {\cal O}(\alpha^4) \, , \\
\label{3rd}
&&V^{\prime \prime \prime}(\phi_0) = 32\frac{m_{\phi}^2}{\phi_0} + {\cal O}(\alpha^2) \, .
\end{eqnarray}
From now on we only keep the leading order terms in all expressions. Note that in gravity-mediated SUSY breaking, the $A$-term and the soft SUSY breaking mass are of the same order of magnitude as the gravitino mass, i.e. $m_{\phi} \sim A \sim m_{3/2} \sim (100~{\rm GeV}-1~{\rm TeV})$. Therefore the condition in Eq.~(\ref{dev}) can indeed be satisfied. We then have $\phi_0 \sim {\cal O}(10^{14}~{\rm GeV})$.

Inflation occurs within an interval
\beq \label{plateau}
\vert \phi - \phi_0 \vert \sim {\phi^3_0 \over 60 M^2_{\rm P}} ,
\eeq
in the vicinity of the point of inflection, within which the slow roll parameters $\epsilon \equiv (M^2_{\rm P}/2)(V^{\prime}/V)^2$ and $\eta \equiv M^2_{\rm P}(V^{\prime \prime}/V)$  are smaller than $1$. The Hubble expansion rate during inflation is given by
\beq \label{hubble}
H_{\rm MSSM} \simeq \frac{1}{\sqrt{45}}\frac{m_{\phi}\phi_0}{M_{\rm P}}
\sim (100~{\rm MeV}-1~{\rm GeV})\,.
\eeq
The amplitude of density perturbations $\delta_H$ and the scalar spectral index $n_s$ are given by~\cite{LK,AEGJM}:
\beq \label{ampl}
\delta_H = {8 \over \sqrt{5} \pi} {m_{\phi} M_{\rm P} \over \phi^2_0}{1 \over \Delta^2}
~ {\rm sin}^2 [{\cal N}_{\rm COBE}\sqrt{\Delta^2}]\,, \eeq
and
\beq \label{tilt}
n_s = 1 - 4 \sqrt{\Delta^2} ~ {\rm cot} [{\cal N}_{\rm COBE}\sqrt{\Delta^2}], \eeq
respectively where
\beq \label{Delta}
\Delta^2 \equiv 900 \alpha^2 {\cal
N}^{-2}_{\rm COBE} \Big({M_{\rm P} \over \phi_0}\Big)^4\,. \eeq
${\cal N}_{\rm COBE}$ is the number of e-foldings between the time when the observationally relevant perturbations are generated till the end of inflation and follows: ${\cal
N}_{\rm COBE} \simeq 66.9 + (1/4) {\rm ln}({V(\phi_0)/ M^4_{\rm P}}) \sim 50$, provided that the universe is immediately thermalized after the end of inflation~\cite{MULTI}.
We note that reheating after MSSM inflation is very fast, due to gauge couplings of the inflaton to gauge/gaugino fields, and results in a radiation-dominated universe within few Hubble times after the end of inflation~\cite{AEGJM}.

\subsection{Parameter space of MSSM inflation}

\begin{figure}
\vspace*{-0.0cm}
\begin{center}
\includegraphics[width=7.0cm]{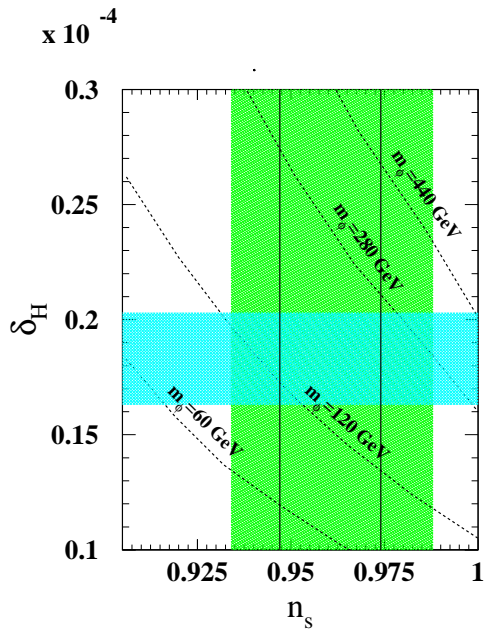}
\caption{$n_s$ is plotted as a function of $\delta_H$ for different
values of $m_{\phi}$. The $2\sigma$ region for $\delta_H$ is shown by the blue horizontal band and the $2\sigma$ allowed region of $n_s$ is shown by the vertical green band. The $1\sigma$ allowed region of $n_s$ is within the solid vertical lines. We choose
$\lambda =1$.} \label{nsdel0}
\end{center}
\end{figure}


A remarkable property of MSSM inflation, which is due to inflation occurring near a point of inflection, is that it can give rise to a wide range of scalar spectral index. This is in clear distinction with other models (for example, chaotic inflation, hybrid inflation, natural inflation, etc.~\cite{Lyth}) and makes the model very robust. Indeed it can yield a spectral index within the whole $2 \sigma$ allowed range by 5-year WMAP data $0.934 \leq n_s \leq 0.988$~\footnote{Note that for for $\alpha^2 = 0$, Eqs.~(\ref{ampl},\ref{tilt}) are
reduced to the case of a saddle point inflation, for which the spectral index
is strictly $0.92$, for details see~\cite{AEGM}.
For $\alpha^2 < 0$, the spectral index will be smaller than the $0.92$, which is more than $3 \sigma$ away
from observations. The more interesting case, as pointed out
in~\cite{AEGJM}, happens for $\alpha^2 > 0$.}. This happens for
\beq \label{Delta2}
2 \times 10^{-6} \leq \Delta^2 \leq 5.2 \times 10^{-6}\,. \eeq
In Fig.~(\ref{nsdel0}), we show $\delta_H$ as a function of $n_s$ for
different values of $m_{\phi}$. The horizontal blue band shows the $2 \sigma$ experimental band for $\delta_H$. The vertical green shaded region is the 2$\sigma$ experimental band for $n_s$. The region enclosed by  solid lines shows the 1$\sigma$ experimental allowed region.

We find that smaller values of $m_{\phi}$ are preferred for smaller values of $n_s$. We also find
that the allowed range of $m_{\phi}$ is $90-330$~GeV for the
experimental ranges of $n_s$ and $\delta_H$. This figure is drawn for $\lambda \simeq
1$, which is natural in the context of effective field theory (unless it is suppressed due to some symmetry). Smaller values of $\lambda$ will lead to an increase in $m_{\phi}$~\cite{ADM}.


\section{False vacuum inflation}

It is conceivable that the universe gets trapped in a metastable vacuum at some stage irrespective of how it began~\footnote{A very attractive scenario to conceive our universe is through a bouncing cosmology~\cite{Bouncing} or cyclic universe~\cite{Cyclic}. In all these proposals it is still possible to realize a false vacuum inflation at later times, as the existence of a false vacuum is common to a filed content and an effective field theory. Therefore, neither cyclic nor bouncing cosmology is in contradiction with a (later) stage false vacuum inflation. There are also proposals of creating perturbations through closed strings~\cite{Brandenberger} during the Hagedorn phase of strings.}, or whether it went though an early phase of inflation at high scales~\footnote{Foremost there are plethora of models of inflation inspired from string theory~\cite{Kallosh} and particle physics ~\cite{Lyth}. However, in the former reheating to the observable sector remains a challenging task~\cite{Myers}, while the latter generically invokes an {\it absolute gauge singlets} to drive inflation with the couplings and masses put by hand to match observations.}.

Once the energy density of the false vacuum dominates, inflation begins and the universe undergoes accelerated expansion with a constant Hubble rate $H_{\rm false}$. False vacuum is not stable and decays via bubble nucleation. The rate per volume for the decay of a metastable vacuum to the true
vacuum takes the form
\beq \label{tunnel} \Gamma/V = C \exp\left(-\Delta S_E \right)\ ,\eeq
where $C$ is a one-loop determinant and $\Delta S_E$ is the difference in Euclidean actions between the instanton and the
background with larger cosmological constant. The determinant $C$ can at most be $C\lesssim M^4_{\rm P}$, simply because $M_{\rm P}$
is the largest scale available, and estimates (ignoring metric fluctuations) give a value as small as $C\sim r^{-4}_0$, with $r_0$
the instanton bubble radius \cite{Instanton,hep-ph/9308280}. Therefore if we look at a decay rate in a (comoving) Hubble volume, we find
\beq \label{tunnel2}
\Gamma \lesssim \frac{M^4_{\rm P}}{H^3_{\rm false}}\exp\left(-\Delta S_E \right)\,. \eeq
Especially with a large Hubble scale $H_{\rm false}$, the associated
decay time is much longer than $H^{-1}_{\rm false}$, given that
typically $\Delta S_E \gg 1$. This implies that most of the space is
locked in the false vacuum and inflate forever, while bubble
nucleation creates pockets of true vacuum whose size grow only linearly
in time.


\subsection{Quantum Fluctuations of MSSM inflaton}

During false vacuum inflation quantum fluctuations displace any scalar field whose mass is smaller than $H_{\rm false}$. The question is whether these
fluctuations can push the MSSM inflaton sufficiently to the plateau
of its potential around the point of inflection $\phi_0$, see Eqs.~(\ref{infvev},\ref{plateau}).
If MSSM inflaton begins with a small VEV $\phi < \phi_0$, the mass term in Eq.~(\ref{scpot}) dominates.
Hence, for any $H_{\rm false} > {\cal O}({\rm TeV})$, it
obtains a quantum jump, induced by the false vacuum inflation, of
length $H_{\rm false}/2\pi$, within each Hubble
time~\cite{LINDE}~\footnote{To be more precise, the quantum
fluctuations have a Gaussian distribution, and the r.m.s
(root mean square) value of the modes which exit the inflationary
Hubble patch within one Hubble time is given by $H_{\rm
false}/2\pi$.}. These jumps superimpose in a random walk fashion,
which is eventually counterbalanced by the classical slow roll due to
the mass term, resulting in~\cite{LINDE}
\beq \label{fluct} \langle \phi^2 \rangle = {3 H^4_{\rm false} \over
8 \pi^2 m^2_{\phi}} \Big[1 - {\rm exp}\Big(-{2 m^2_{\phi} \over 3
H_{\rm false}} t\Big)\Big], \eeq
which for $t \to \infty$ yields
\beq \label{rms}
\phi_{r.m.s} = \sqrt{3 \over 8 \pi^2} {H^2_{\rm
false} \over m_{\phi}}\,.
\eeq

If $\phi_{r.m.s} \geq \phi_0$, then $\phi$ will lie near
$\phi_0$, see Eq.~(\ref{infvev}),in many regions of space. This requires that
\beq \label{cond1}
H_{\rm false} \geq \Big({8 \pi^2 \over 3}\Big)^{1/4} (m_{\phi} \phi_0)^{1/2} \gs 10^8~{\rm GeV}.
\eeq
$\phi_{r.m.s}$ settles at its final value when $t > 3 H_{\rm false}/2m_{\phi}^2$, which
amounts to
\beq \label{flucefold}
N_{\rm false} > {3 \over 2} \Big({H_{\rm false} \over m_{\phi}}\Big)^2 \gs 10^{16},
\eeq
e-foldings of false vacuum inflation. The large number of e-foldings required is
not problematic as inflation in the false vacuum is eternal in nature.


\subsection{Inflection point as a dynamical attractor}

In general the MSSM inflaton can have a VEV above the point of inflection $\phi > \phi_0$ and/or a large velocity ${\dot \phi}$ at the beginning of false vacuum inflation. In this case the classical motion of the field becomes important. This is found from the equation of motion,
\beq \label{eqm}
\ddot \phi + 3 H_{\rm false} {\dot \phi} = -V'(\phi) \,,
\eeq
%
where $V(\phi)$ is given by Eq.~(\ref{scpot}). There are typically three regimes in the evolution of $\phi$ field, which we discuss below.


\subsubsection{Oscillatory regime}

If the initial VEV of $\phi$, denoted by $\phi_i$, is such that $V^{\prime \prime}(\phi_i) > H^2_{\rm false}$, then it starts in the oscillatory regime. It can be seen from Eq.~(\ref{infvev}) that for $\phi \gs \phi_0$ the last term on the right-hand side of Eq.~(\ref{scpot}) dominates $V(\phi)$, i.e. $V(\phi) \propto \phi^{10}$. Hence we are in the oscillatory phase if
\beq \label{osccond}
\phi_{i} \gs \phi_{\rm tr} \sim (H_{\rm false} M^3_{\rm P})^{1/4}.
\eeq
For $V(\phi) \propto \phi^n$, the Hubble expansion redshifts the amplitude of oscillations $\propto a^{-6/n+2}$ (where $a$ is the scale factor of the universe)~\cite{STB}. In the case at hand $n=10$, implying that $\phi \propto a^{-1/2}$, which during false vacuum inflation reads $\phi \propto {\rm exp}(-H_{\rm false}t/2)$. Oscillations end once $\phi$ is redshifted down to the VEV in Eq.~(\ref{osccond}). The maximum time spent in the oscillatory regime (happening if $\phi_i \sim M_{\rm P}$) is found to be
\beq \label{oscdur}
{t}_{\rm osc} < {1 \over 2} H^{-1}_{\rm false} ~ {\rm ln} \Big({M_{\rm P} \over H_{\rm false}}\Big),
\eeq
which amounts to several e-foldings of false vacuum inflation.


\subsubsection{Kinetic energy dominance regime}

If $\phi_i < \phi_{\rm tr}$, then $V^{\prime \prime}(\phi) < H^2_{\rm false}$, and the potential is flat during false vacuum inflation. The dynamics of $\phi$ in this case depends on its initial velocity denoted by ${\dot \phi}_i$. If ${\dot \phi}^2_i > 2 V(\phi_i)$, we are in the kinetic energy dominance phase where $\rho_{\phi} \approx {\dot \phi}^2/2$, which implies that ${\ddot \phi} \approx - 3 H_{\rm false} {\dot \phi}$. Hence in this regime we have ${\dot \phi} \propto {\rm exp} (-3 H_{\rm false} t)$
which results in
\beq \label{kin}
\phi - \phi_i = {1 - {\rm exp}[-3 H_{\rm false} (t - t_i)] \over 3 H_{\rm false}} {\dot \phi}_i .
\eeq
%
It is seen that the maximum distance that $\phi$ can move during this phase is ${\dot \phi}/3H_{\rm false}$. Hence, if $\phi_i > \phi_0$, the field will end up at $\phi > \phi_0$, provided that
%
$\phi_i - \phi_0 > -{{\dot \phi}_i/3 H_{\rm false}}$.
%
This implies that the inflaton will overshoot the point of inflection only if it begins very close to $\phi_0$ and has a large negative velocity initially. It is interesting to note that $\phi$ can end up above the point of inflection even if $\phi_i < \phi_0$, provided that ${\dot \phi}_i > 3 H_{\rm false} (\phi_0 -\phi_i)$.

The kinetic energy dominance regime ends when ${\dot \phi}^2 \sim 2 V(\phi)$. Thus, since ${\dot \phi} \propto {\rm exp}(-3 H_{\rm false} t)$, the duration of this phase is given by
\beq \label{kindur}
{t}_{\rm kin} \ls {1 \over 6} H^{-1}_{\rm false} ~ {\rm ln}\Big({{\dot \phi}^2_i \over 2 V(\phi_0)} \Big),
\eeq
which is typically a few e-foldings of false vacuum inflation.

\subsubsection{Slow roll regime}

Once an initial phase of oscillations or kinetic energy dominance ends, $\phi$ starts a slow roll motion towards $\phi_0$~\footnote{It will be in the slow roll regime from the beginning if $\phi_i < \phi_{\rm ytr}$ and ${\dot \phi}^2_i < V(\phi_0)$.}. The equation of motion in this regime is $3H_{\rm false} \dot \phi + V^{\prime}(\phi)\approx 0$. Initially the field is under the influence of the non-renormalizable potential, i.e. $V(\phi) \propto \phi^{10}$, for which
\beq
\frac{\phi}{\phi_i} \simeq \left(1+\frac{4}{(15)}\frac{V^{\prime \prime}(\phi_i)}{H_{\rm false}} t \right)^{-\frac{1}{8}}  \, .
\eeq
This part of motion takes a maximum time 
\begin{eqnarray} \label{non}
t_{\rm non} \sim 2 H^{-1}_{\rm false} \left({H^2_{\rm false} \over V^{\prime \prime}(\phi_i)}\right) \Big({\phi_{\rm tr} \over \phi_0}\Big)^8 \sim 2 H^{-1}_{\rm false} \Big({H_{\rm false} \over m_{\phi}}\Big)^2 \, , \nonumber \\
\, \nonumber \\
\end{eqnarray}
which amounts to $\sim (H_{\rm false}/m_{\phi})^2$ e-foldings of false vacuum inflation.

As $\phi$ moves toward $\phi_0$, the $\phi^{10}$ term becomes increasingly smaller. Eventually, for $\phi \approx \phi_0$, we have $V^{\prime}(\phi) = V^{\prime}(\phi_0) + V^{\prime \prime \prime}(\phi_0) (\phi - \phi_0)^2/2$. As we will see, this is the longest part of $\phi$ journey. We would like to find how long does it take for $\phi$ to reach the edge of the plateau in Eq.~(\ref{plateau}). Outside the plateau, $V^{\prime}(\phi_0)$ is subdominant, see Eqs.~(\ref{1st},~\ref{3rd}), and hence we obtain
\beq \label{slow}
{\dot \phi} = - {32 m^2_{\phi} (\phi - \phi_0)^2 \over 3 H_{\rm false} \phi_0}.
\eeq
This results in
\beq \label{att}
(\phi - \phi_0) \approx {3 H_{\rm false} \phi_0 \over 32 m^2_{\phi} t},
\eeq
for large $t$. Therefore the inflection point acts as an attractor for the classical equation of motion. After using Eq.~(\ref{plateau}), we find the time that it takes for $\phi$ to reach the edge of the plateau~\footnote{Actually this will be faster due to effect of quantum fluctuations. As $\phi$ gets closer to the inflection point, $V^{\prime}(\phi)$ becomes smaller. Once $V^{\prime}(\phi) \sim 3 H^3_{\rm false}/2 \pi$, quantum jumps dominate the slow roll motion. This happens at $(\phi - \phi_0) \sim (3H^3_{\rm false} \phi_0/32 \pi m^2_{\phi})^{1/2}$. From this point on fluctuations take over and move $\phi$ to the plateau in random walk fashion.}
\beq \label{slowdur}
t_{\rm slow} \gs 180 H^{-1}_{\rm false} \Big({H_{\rm false} \over m_{\phi}}\Big)^2 \Big({M_{\rm P} \over \phi_0}\Big)^2,
\eeq
which amounts to $\gs 10^{10}$ e-foldings of false vacuum inflation (note that $\phi_0 \sim 10^{14}$ GeV, and $H_{\rm false} > m_{\phi}$). It is seen from Eqs.~(\ref{oscdur},\ref{kindur},\ref{non},\ref{slowdur}) that the total time that it takes for $\phi$ to settle within the plateau is dictated by the slow roll regime. Hence the total number of e-foldings required to achieve this is
given by
\beq \label{attefold}
N_{\rm false} \gs 180 \Big({H_{\rm false} \over m_{\phi}}\Big)^2 \Big({M_{\rm P} \over \phi_0}\Big)^2 > 10^{10}.
\eeq
After this time $\phi$ is settled within the plateau in the bulk of the inflating space. This implies that the bubbles which nucleate henceforth have the right initial conditions for a subsequent stage of MSSM inflation.

\begin{figure}
\vspace*{-0.0cm}
\begin{center}
\includegraphics[width=8.5cm]{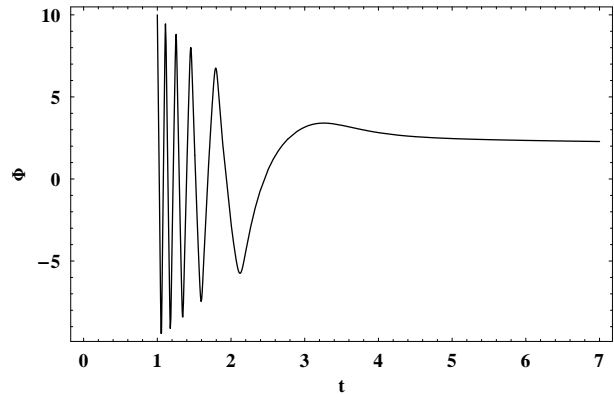}
\end{center}
\caption{We plot $\phi$ (scaled by $\phi_0$) as a function of time for $\phi_i = 10$ and ${\dot \phi}_i = 100$, with $H_{\rm false} = 10^3 m_{\phi}$. There is an initial oscillatory phase since $V^{\prime\prime}(\phi_i) \geq H^2_{\rm false}$. It ends quickly as the amplitude of oscillations decreases fast due to Hubble expansion. Then the slow roll motion begins which lasts much longer.}
\label{oscillations}
\end{figure}

\begin{figure}
\vspace*{-0.0cm}
\begin{center}
\includegraphics[width=8.5cm]{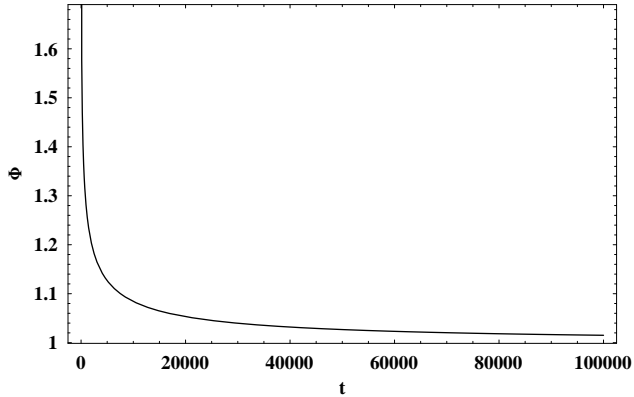}
\end{center}
\caption{The slow roll phase for the same initial conditions as in Fig. (2). It lasts very long but the field asymptotes to the {\it point of inflection} $\phi_0$ with $\dot \phi\rightarrow 0$. }
\label{slowroll}
\end{figure}

\begin{figure}
\vspace*{-0.0cm}
\begin{center}
\includegraphics[width=8.5cm]{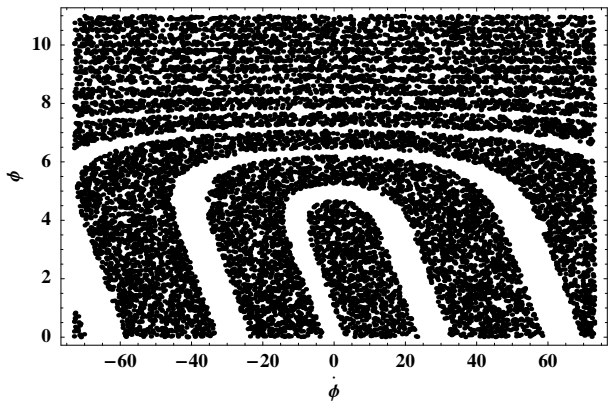}
\end{center}
\caption{We plot initial values of $\phi$ versus $\dot\phi$ for $H_{\rm false}=10^2 m_{\phi}$. The dots show the initial values for which $\phi$ settles to $\pm\phi_0$ and the white bands
$\cap$ show the critically damped regions where $\phi$ settles to zero.}
\label{phase1}
\end{figure}

\begin{figure}
\vspace*{-0.0cm}
\begin{center}
\includegraphics[width=8.5cm]{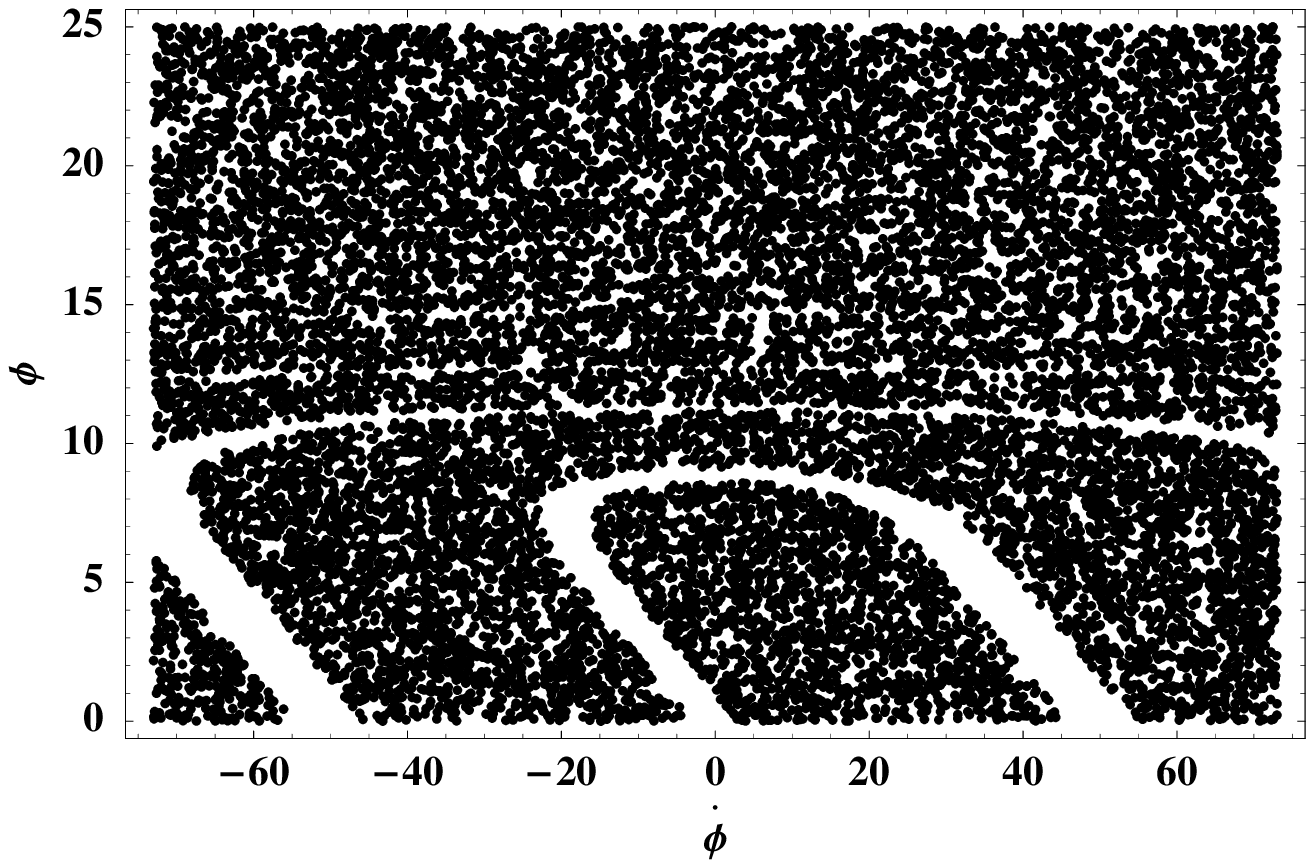}
\end{center}
\caption{Same as in Fig.~(\ref{phase1}) for $H_{\rm false}=10^3 m_{\phi}$.}
\label{phase2}
\end{figure}

\begin{figure}
\vspace*{-0.0cm}
\begin{center}
\includegraphics[width=8.5cm]{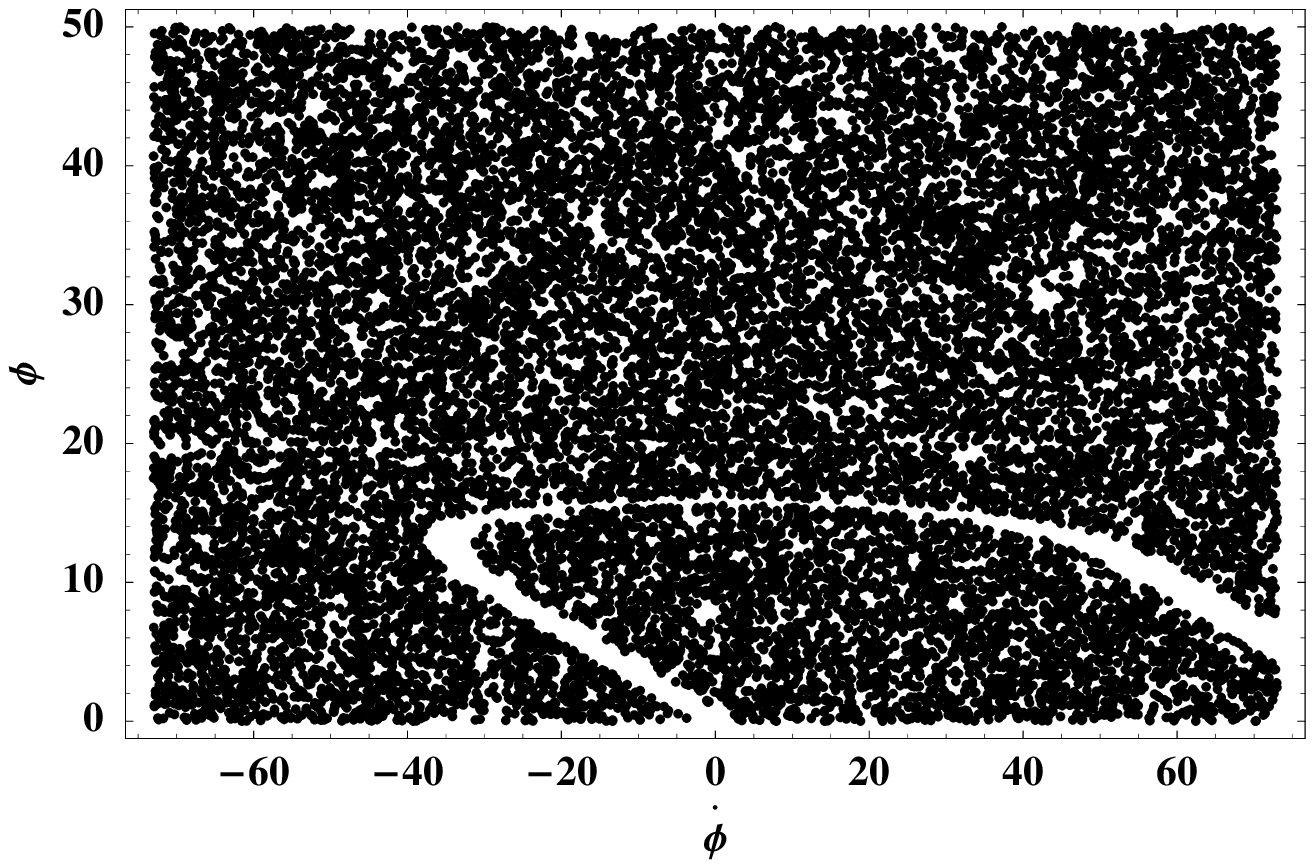}
\end{center}
\caption{Same as in Fig.~(\ref{phase1}) for $H_{\rm false}=10^4 m_{\phi}$.}
\label{phase3}
\end{figure}

\begin{figure}
\vspace*{-0.0cm}
\begin{center}
\includegraphics[width=7.5cm]{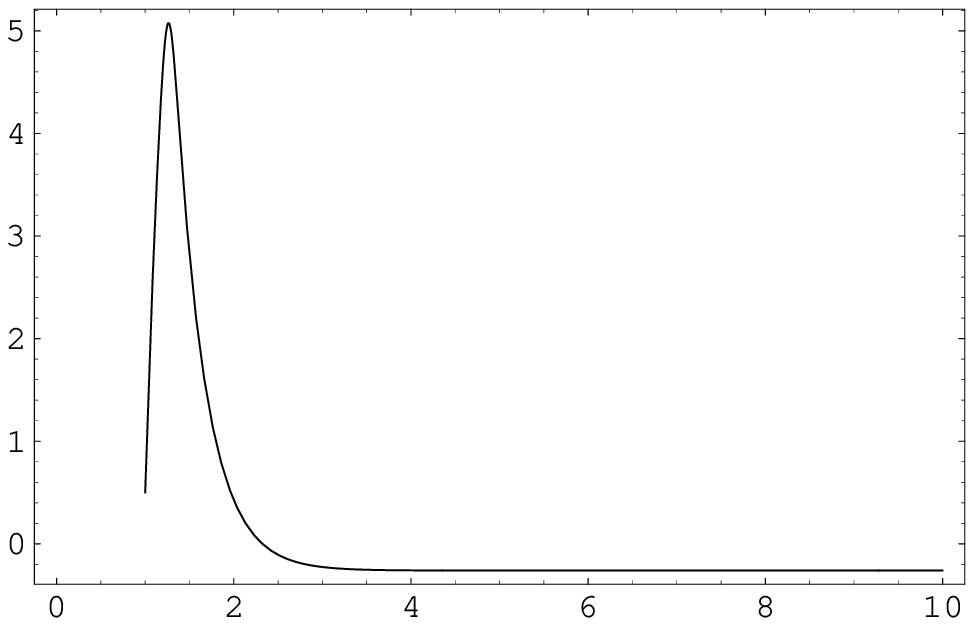}\\
\includegraphics[width=8.0cm]{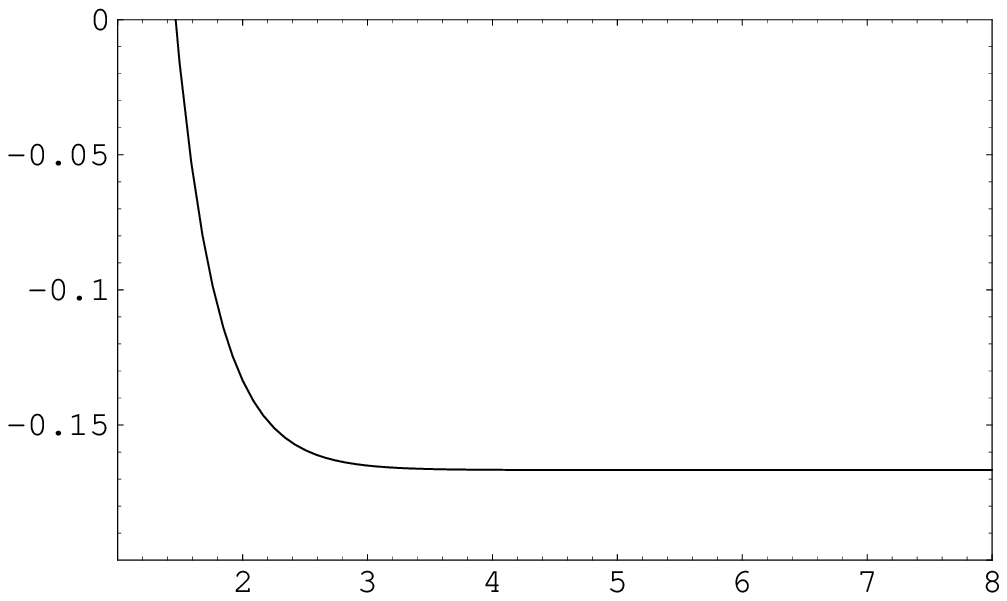}
\end{center}
\caption{We plot $\phi$ as a function of time when $\phi$ does not settle to $\phi_0$ for two sets of initial conditions:
$\phi_i=0.5$, $\dot\phi_i=28$ (top panel) and $\phi_i=0.5$,
$\dot\phi_i=-2$ (bottom panel). We have chosen $H_{\rm false} = 10^2 m_{\phi}$. These are the typical trajectories for $\cap$ shaped bands shown in
Figs.~(\ref{phase1},\ref{phase2},\ref{phase3}).}
\label{critical}
\end{figure}

\subsection{Numerical Results}

Figs.~(\ref{oscillations},\ref{slowroll}) show the evolution of $\phi$ for initial conditions $\phi_i = 10$ and ${\dot \phi}_i = 100$. We have chosen $H_{\rm false} = 10^3 m_\phi$. We have scaled $\phi$ by $\phi_0 \sim 10^{14}$ GeV, see Eq.~(\ref{infvev}), ${\dot \phi}$ by $H_{\rm false} \phi_0$ and time by $H^{-1}_{\rm false}$. Fig~(\ref{oscillations}) shows the initial oscillatory phase (note that $\phi_i > \phi_{\rm tr}$). Oscillations end within few Hubble times as discussed before, see Eq.~(\ref{oscdur}).
Fig.~(\ref{slowroll}) shows the subsequent slow roll motion of $\phi$ towards $\phi_0$, during which $\phi \rightarrow \phi_0$ and ${\dot \phi} \rightarrow 0$. We see that, as explained before, the slow roll phase takes a much longer time than the oscillatory phase. The attractor behavior holds for larger values of $H_{\rm false}$, but it takes longer for $\phi$ to settle within the plateau of the potential around $\phi_0$, in agreement with Eq.~(\ref{attefold}).

The attractor behavior holds for a wide range of initial values of $\phi$ and ${\dot \phi}$. This
feature is evident in Figs.~(\ref{phase1},\ref{phase2},\ref{phase3}), where we have randomly generated $20,000$ points for $H = 10^2 m_{\phi}$, $10^3 m_{\phi}, 10^4 m_{\phi}$ respectively. The black dots are generated when $\phi$ settles to $\phi_0$ or -$\phi_0$. For Fig.~(\ref{phase1}), about $70\%$ of the generated points are in the dots. The situation gets better for larger values of $H_{\rm false}$: in Figs.~(\ref{phase2},\ref{phase3}) the number of dots are about $85\%$ and $90\%$ of the generated points respectively. Note that in large parts of the phase space $\phi$ begins in the oscillatory phase.

Points inside the $\cap$ shaped white bands show the initial conditions for which $\phi$ does not settle at $\phi_0$ or ($-\phi_0$) at late times~\footnote{The plots are symmetric under $\phi \rightarrow -\phi$. Here we have shown the upper half of $\phi-{\dot \phi}$ plane where $\phi \geq 0$.}. Here we focus on the inner most band as the outer ones correspond to initial conditions for which the field ends up being in the inner most band in the course of its motion. The left leg of the band starts at the origin which is expected: as long as $\phi_i < \phi_0$ and ${\dot \phi}_i = 0$, $\phi$ will never make it to the point of inflection. The leg is tilted to the left (${\dot \phi} <$ 0) instead of being vertical, which can be understood as follows. For $\phi_i > \phi_0$ the field  eventually settles at $\phi_0$ unless it has a large negative velocity. In this case it starts in the kinetic dominance regime and overshoots $\phi_0$ before the slow roll regime begins. However, if the negative velocity becomes very large, the field will also overshoot $-\phi_0$ on the other side before the slow roll motion begins. In this case it will eventually settle at $-\phi_0$, which is also a point of inflection and equally good for MSSM inflation.

As we move to larger $\phi$ the left leg turns into a horizontal band. This signals that $\phi_i$ approaches $\phi_{\rm tr}$. Since $V^{\prime \prime}(\phi_{\rm tr}) \sim H^2_{\rm false}$, the motion is critically damped around this point. This implies a fast roll motion, and hence $\phi$ will overshoot $\phi_0$ even if its initial velocity ${\dot \phi}_i$ is zero. In fact the field can get close to $\phi_{\rm tr}$ and undergo fast roll motion even if $\phi_i$ is small, provided that the initial velocity ${\dot \phi}_i$ is large and positive. The right leg of the $\cap$ shaped band shows all of the points for which this happens. Note that the smaller $\phi_i$, the larger is required ${\dot \phi}_i$ in order for $\phi$ to get close to $\phi_{\rm tr}$. This is why the right leg is also tilted to the left. If ${\dot \phi}_i$ becomes larger, the field will go beyond $\phi_{\rm tr}$. This will lead to an oscillatory motion of $\phi$. As explained, it ends within a few Hubble times and is followed by slow roll motion towards $\phi_0$ or $-\phi_0$.

One feature in Figs.~(\ref{phase1},\ref{phase2},\ref{phase3}) is that
the $\cap$ shape fattens (i.e. the distance between the left and right legs gets larger) as $H_{\rm false}$ increases~\footnote{At the first glance, it may seem that the $\cap$ shape gets further tilted to the left and the horizontal part gets thinner as $H_{\rm false}$ increases. However, these are not real effects as the vertical axis is drawn for different ranges of $\phi$ in different figures.}. The reason being that $\phi_{\rm tr} \sim (H_{\rm false} M_{\rm P})^{1/4}$ and $V(\phi_{\rm tr}) \propto \phi^{10}_{\rm tr}$, and hence the initial velocity needed to get $\phi$ past $\phi_{\rm tr}$ grows $\propto H^{5/4}_{\rm false}$ as a function of $H_{\rm false}$. Then, since ${\dot \phi}$ is scaled $\propto H_{\rm false}$ in the figures, the $\cap$ shape fattens $\propto H^{1/4}_{\rm false}$. In fact, this is the reason why the attractor behavior holds for a wider range of initial conditions as $H_{\rm false}$ increases.

In Fig.~(\ref{critical}) we show a couple of examples of field motion for initial conditions which fall within the left and the right legs of the inner most $\cap$ for $H_{\rm false} = 10^2 m_\phi$. In the first case (top panel) $\phi$ moves fast towards $\phi_{\rm tr}$, then undergoes fast roll motion, as a result of which it overshoots $\phi_0$, and eventually settles at the origin. In the second case (bottom panel) the field overshoots $\phi_0$ while being in the kinetic energy dominance phase, and then undergoes slow roll motion towards the origin. These examples represent typical trajectories for the $\cap$ shaped bands.

\section{Dynamics inside the bubble and MSSM inflation}

Here we consider the bubbles that have the right initial conditions for MSSM inflation, i.e. $\phi$ has settled in the plateau around a point of inflection according to Eq.~(\ref{plateau}). The initial size of the bubble is $r_0 < H^{-1}_{\rm false}$. Inside of an expanding bubble has the same geometry as an open FRW universe. The Hubble rate inside the bubble is therefore given by
\beq \label{bubhub} H^2 = {V_{\phi} + V_{\varphi} \over 3 M^2_{\rm P}} + {1
\over a^2}, \eeq
where $a$ is the scale factor of the universe and $V_{\varphi}$ is the energy density in the $\varphi$ field.
Note that $\varphi$ is the field responsible for forming the false vacuum, which could either arise within MSSM or from some other sector. In the last section we will provide an example of $\varphi$ field as an MSSM flat direction.


Since $\phi$ is inside the plateau of its potential, its dynamics is frozen, hence $V(\phi)\sim V(\phi_0)$ as long as $H > H_{\rm MSSM}$. Right after tunneling, $H \equiv {\dot a}/a = r^{-1}_0 > H_{\rm false}$. This implies that the last term on the right-hand side of Eq.~(\ref{bubhub}) dominates over the first two terms, and hence the
universe is curvature dominated. The $\varphi$ field oscillates around the
true vacuum of its potential at the origin, and quickly decays to radiation whose energy density is redshifted $\propto a^{-4}$. On the other hand the curvature term is redshifted $\propto a^{-2}$, while $V(\phi)$ remains essentially constant (due to extreme flatness of the inflaton potential) for $H > H_{\rm MSSM}$. As a result, the universe inside the bubble will remain curvature dominated until $H \simeq H_{\rm MSSM}$.

At this point the inflaton field $\phi$ dominates the energy density and a phase of MSSM inflation begins. This blows the open universe inside the bubble and inflates away the curvature term. As long as the total number of e-foldings is $N_{\rm COBE}$ plus few, the observable part of the universe looks like flat today (within the limits of 5 year WMAP data)~\cite{WMAP}. Perturbations of the correct size with acceptable spectral index will be generated during the slow roll phase, and the SM degrees of freedom will be created from the decay of $\phi$ field in the post-inflationary phase~\cite{AEGJM}.


\section{Metastable vacua in MSSM}

It is interesting that MSSM itself may have metastable vacua that can drive a phase of false vacuum inflation discussed above. Here we present a simple example when such a vacuum may arise.

Let us begin by introducing the MSSM superpotential:
\beq \label{mssmsup} W_{\rm MSSM} = y_u Q H_u u + y_d Q H_d d + y_e
L H_d e + \mu H_u H_d, \eeq
where $Q$, $L$ are the left-handed quark and lepton superfields respectively, $u,~d,~e$ are the right-handed up- and down-type quark and lepton superfields respectively, and $H_u,~H_d$ are the Higgs doublets. Here $y_u,~y_d,~y_e$ are Yukawa couplings of the
up-type quarks, down-type quarks, and charged leptons respectively.
The parameter $\mu \sim {\cal O}({\rm TeV})$ is required for an
acceptable electroweak symmetry breaking. Here $y_u,~y_d,~y_e$ are
$3 \times 3$ matrices and we have omitted the family indices on the
quarks and leptons for simplicity. We work in the basis where the
Yukawa couplings are diagonalized. Each of the monomials $Q H_u u$,
$Q H_d d$, $Q H_d e$ represent a D-flat direction. Note that the
MSSM Yukawa couplings lift the F-flatness at the renormalizable
($n=3$) superpotential level. However, as we will see, higher order
terms result in a non-trivial modification to the MSSM potential.

Let us consider $Q H_u u$ monomial (with associated Yukawa coupling $y Q H_u u$).
It represents a flat direction
%
\beq \label{flat}
\varphi = \frac{Q + H_u + u}{\sqrt{3}} \,.
\eeq
%

Under general circumstances, allowed by gauge invariance, the
superpotential also includes higher order terms of the form $(Q H_u u)^n$ and $(QH_u u)^n (H_u H_d)$. After rewriting them in terms of $\varphi$ and $\chi \equiv H_d$, the superpotential reads (to the lowest order)
%
%
%
\beq \label{nonren1}
W \supset y {\varphi^3 \over 3} + \lambda
{\varphi^6 \over 6 M^3_{\rm P}} + \lambda^{\prime} {\varphi^4 \chi \over 4
M^2_{\rm P}} + ... ,
\eeq
where $\lambda,~\lambda^{\prime}$ are constants of ${\cal O}(1)$. The scalar potential along the flat direction follows:
\begin{eqnarray} \label{pot1}
V(\varphi) = \vert y \varphi^2 + \lambda
{\varphi^5 \over M^3_{\rm P}}\vert^2 +
\Big({{\lambda^{\prime}}^2 \over 16} {\vert \varphi \vert^8 \over
M^4} \Big) + ... \, .
\end{eqnarray}
%
We also have the contribution from soft supersymmetry breaking terms $V_{\rm
soft}$. However, as we will see shortly, these terms are subdominant.

The first term on the right-hand side of Eq.~(\ref{pot1}) vanishes at the origin $\varphi = 0$ and three
other points with radial and angular coordinates $(\varphi_0,\theta_0)$
\beq \label{false}
\varphi_0 = \Big({y \over \lambda}\Big)^{1/3} M
~~,~~ 3 \theta_0 + \theta_{\lambda} - \theta_y = \pi,~3\pi,~5\pi.
\eeq
Here $\theta_y,~\theta_{\lambda}$ are the phases of $y$ and
$\lambda$ couplings respectively. Note that $y \ll 1$ except for the
the top quark Yukawa, and hence $\varphi_0 \ll M_{\rm P}$. This implies that the higher order terms in Eq.~(\ref{nonren1}), depicted as ``..'', have a subleading contribution and are indeed negligible.

The minima in Eq.~(\ref{false}) are separated from the origin by a barrier whose height is given by
\beq \label{barrier}
V_{\rm barrier} = {9 \over 25} \Big({2 \over 5}\Big)^{4/3} y^2 \Big({y \over \lambda}\Big)^{4/3} M^4_{\rm P}.
\eeq
The second term on the right-hand side of Eq.~(\ref{pot1}) is positive
definite and lifts the potential at $\vert \varphi \vert \neq 0$. This
results in having three false minima at $\vert \varphi \vert \sim
\varphi_0$. Depending on the relative size of
$y,~\lambda,~\lambda^{\prime}$, the potential at the false minima
$V_{\rm false}$ can assume any value
\beq \label{flasepot} V_{\rm false} \ls V_{\rm barrier}. \eeq
%
False vacuum inflation in these minima has a Hubble expansion rate:
\beq \label{Hubble} H_{\rm false} \ls \frac{V_{\rm
barrier}^{1/2}}{\sqrt{3} M_{\rm P}} \ls y^{5/3} M_{\rm P}\,.
\eeq
For $y \sim 10^{-5}-10^{-2}$ (which is the case for all of the
SM Yukawa couplings except that of the top quark) the false vacuum inflation could
be driven at a Hubble rate as large as $H_{\rm false} \sim
10^{14}$~GeV. Note that, as mentioned earlier, the contribution of the supersymmetry breaking terms to the false vacuum potential is
negligible for $H_{\rm false} \gg {\cal O}({\rm TeV})$.



\section{Discussion and Conclusion}

We discussed how the observable patch of the universe could emerge after bubble nucleation
from a false vacuum via MSSM inflation. We discussed how the initial conditions for MSSM inflation can be set in the background of a false vacuum dominated inflation. In particular, we showed in detail that a large phase space is available where the trajectories asymptote towards the {\it point of inflection} around which MSSM inflation takes place, thus making it a dynamically attractor solution. We also furnished a simple example of constructing a false vacuum within MSSM, indicating that both of the false vacuum inflation and a subsequent slow roll inflation could occur within the MSSM.

\vskip30pt

{\it Acknowledgments-} We would like to thank Andrew Frey, Pratika Dayal, Kari Enqvist, Juan Garcia-Bellido, Gordy Kane and Lev Kofman for discussions on various aspects of this project. RA research was supported in part by Perimeter Institute for Theoretical Physics. The work of BD is supported in part by DOE grant DE-FG02-95ER40917, and he would like to thank IPPP (Durham) for their kind hospitality while this work was being completed. The work of AM is partly supported by the European Union through Marie Curie Research and Training Network ``UNIVERSENET'' (MRTN-CT-2006-035863), and he would like to thank KITP (Santa Barbara) and SISSA (Trieste) for their hospitality during which part of the work has been carried out.




\begin{thebibliography}{99}

\bibitem{Douglas}
M.~R.~Douglas,
  Comptes Rendus Physique {\bf 5}, 965 (2004).

\bibitem{Guth}
 A.~H.~Guth,
  Phys.\ Rev.\  D {\bf 23}, 347 (1981).


\bibitem{WMAP}
 E.~Komatsu {\it et al.}  [WMAP Collaboration],
  arXiv:0803.0547 [astro-ph].

\bibitem{BBN}
 K. A. Olive, G. Steigman and T. P. Walker, Phys. Rept. {\bf 333}, 389 (2000).



\bibitem{AEGM}
R.~Allahverdi, K.~Enqvist, J.~Garcia-Bellido and A.~Mazumdar,
  Phys. Rev. Lett. {\bf 97}, 191304 (2006).

\bibitem{AEGJM}
 R.~Allahverdi, K.~Enqvist, J.~Garcia-Bellido, A.~Jokinen and A.~Mazumdar,
  JCAP {\bf 0706}, 019 (2007).

\bibitem{AKM}
R.~Allahverdi, A.~Kusenko and A.~Mazumdar,
  JCAP {\bf 0707}, 023 (2007).

\bibitem{AJM}
 R.~Allahverdi, A.~Jokinen and A.~Mazumdar,
  arXiv:hep-ph/0610243.


\bibitem{MSSM-REV}
 K.~Enqvist and A.~Mazumdar,
  Phys.\ Rept.\  {\bf 380}, 99 (2003).
 M.~Dine and A.~Kusenko,
  Rev.\ Mod.\ Phys.\  {\bf 76}, 1 (2004).

\bibitem{DRT}
M.~Dine, L.~Randall and S.~Thomas, Phys. Rev. Lett. {\bf 75}, 398 (1995).
M. Dine, L. Randall and S. Thomas, Nucl. Phys. B {\bf 458}, 291 (1996).


\bibitem{Curvaton}
 K.~Enqvist, S.~Kasuya and A.~Mazumdar,
  Phys.\ Rev.\ Lett.\  {\bf 90}, 091302 (2003).
K.~Enqvist, A.~Jokinen, S.~Kasuya and A.~Mazumdar,
  Phys.\ Rev.\  D {\bf 68}, 103507 (2003).
M.~Postma and A.~Mazumdar,
  JCAP {\bf 0401}, 005 (2004).
K.~Enqvist, A.~Mazumdar and M.~Postma,
  Phys.\ Rev.\ D {\bf 67}, 121303 (2003).
 R.~Allahverdi, K.~Enqvist, A.~Jokinen and A.~Mazumdar,
  JCAP {\bf 0610}, 007 (2006).

\bibitem{Ours}
R.~Allahverdi and A.~Mazumdar,
  arXiv:hep-ph/0505050.
R.~Allahverdi and A.~Mazumdar,
  JCAP {\bf 0610}, 008 (2006).
R.~Allahverdi and A.~Mazumdar,
  Phys.\ Rev.\ D {\bf 76}, 103526 (2007).
 R.~Allahverdi and A.~Mazumdar,
 JCAP {\bf 0708}, 023 (2007).
R.~Allahverdi and A.~Mazumdar,
  arXiv:0802.4430 [hep-ph], Phys. Rev. D (in press).



\bibitem{Asko}
K.~Enqvist, A.~Jokinen and A.~Mazumdar,
  JCAP {\bf 0411}, 001 (2004).


\bibitem{ADM}
 R.~Allahverdi, B.~Dutta and A.~Mazumdar,
  Phys. Rev. D {\bf 75}, 075018 (2007).


\bibitem{ADM2}
R.~Allahverdi, B.~Dutta and A.~Mazumdar,
  Phys. Rev. Lett. {\bf 99}, 261301 (2007).

\bibitem{LK}
  J.~C.~Bueno Sanchez, K.~Dimopoulos and D.~H.~Lyth,
  JCAP {\bf 0701}, 015 (2007).
   R.~Allahverdi and A.~Mazumdar,
  arXiv:hep-ph/0610069.


\bibitem{AFM}
R.~Allahverdi, A.~R.~Frey and A.~Mazumdar, Phys. Rev. D {\bf 76}, 026001 (2007).


\bibitem{GKM}
 T.~Gherghetta, C.~F.~Kolda and S.~P.~Martin,
  Nucl.\ Phys.\ B {\bf 468}, 37 (1996).


\bibitem{MULTI}
A.~R.~Liddle and S.~M.~Leach,
  Phys.\ Rev.\  D {\bf 68}, 103503 (2003).
 C.~P.~Burgess, R.~Easther, A.~Mazumdar, D.~F.~Mota and T.~Multamaki,
  JHEP {\bf 0505}, 067 (2005).

\bibitem{Bouncing}
T.~Biswas, A.~Mazumdar and W.~Siegel,
  JCAP {\bf 0603}, 009 (2006).


\bibitem{Cyclic}
 P.~J.~Steinhardt and N.~Turok,
  New Astron.\ Rev.\  {\bf 49}, 43 (2005).

\bibitem{Brandenberger}
A.~Nayeri, R.~H.~Brandenberger and C.~Vafa,
  Phys.\ Rev.\ Lett.\  {\bf 97}, 021302 (2006).
 T.~Biswas, R.~Brandenberger, A.~Mazumdar and W.~Siegel,
  JCAP {\bf 0712}, 011 (2007).


\bibitem{Kallosh}
R.~Kallosh and A.~Linde,
  JCAP {\bf 0704}, 017 (2007).



\bibitem{Lyth}
D.~H.~Lyth and A.~Riotto,
  Phys.\ Rept.\  {\bf 314}, 1 (1999).

\bibitem{Myers}
A.~R.~Frey, A.~Mazumdar and R.~C.~Myers,
  Phys.\ Rev.\  D {\bf 73}, 026003 (2006).


\bibitem{Instanton}
R.~Coleman and F.~De Luccia,
  Phys.\ Rev.\ D {\bf 21}, 3305 (1980).


\bibitem{hep-ph/9308280}
J.~Garriga,
  Phys.\ Rev.\  D {\bf 49}, 6327 (1994).

\bibitem{LINDE}
 A.~D.~Linde,
  ``Particle Physics and Inflationary Cosmology,''
  Harwood Academic Publishers, (1993).

\bibitem{STB}
Y.~Shtanov, J.~H.~Traschen and R.~H.~Brandenberger, Phys. Rev. D {\bf 51}, 5438 (1995).






\end{thebibliography}
\end{document}